\documentclass[aps,prc,onecolumn,superscriptaddress]{revtex4}
\usepackage[latin1]{inputenc}
\usepackage{bm}
\usepackage{epsfig}
\usepackage{amsthm}
\usepackage{amsfonts}
\usepackage{float}
\usepackage{amsmath,amssymb}
\usepackage{color}
\usepackage{slashed}
\usepackage{relsize}
\usepackage[toc,page]{appendix} 
\usepackage{wrapfig}
\usepackage{graphicx}
\usepackage{dcolumn}
\usepackage{bm}
\newcommand{\be}{\begin{equation}} 
\newcommand{\ee}{\end{equation}} 
\newcommand{\bea}{\begin{eqnarray}} 
\newcommand{\eea}{\end{eqnarray}} 
\newcommand{\nn}{\nonumber} 
\newcommand{\mintedim}[2]{{\int\kern-0.50em\mbox{{\small$\mathop{\frac{\mbox{{\small${\rm d^{#2}}\vect{#1}$}}}{\mbox{{\small$(2\pi)^{#2}$}}}}$}}\ }} 
\newcommand{\inteonedim}[1]{{\int_0^\infty\kern-1em\mbox{{\small${\rm d}{#1}$}}}} 
\newcommand{\vect}[1]{\bm{#1}} 

\begin{document}
\title{
\textcolor{black} {Jet Quenching of the Heavy Quarks in the Quark-Gluon Plasma \\
and the Nonadditive Statistics}
}

\author{Trambak Bhattacharyya}
\email{trambak.bhattacharyya@gmail.com}
\affiliation{Bogoliubov Laboratory of Theoretical Physics, Joint Institute for Nuclear Research, Dubna, 141980, \\
Moscow Region, Russia}

\author{Eugenio Meg\'{i}as}
\email{emegias@ugr.es}
\affiliation{Departamento de Física At\'{o}mica, Molecular y Nuclear and Instituto Carlos I de Física Te\'{o}rica y Computacional, Universidad de Granada, Avenida de Fuente Nueva s/n, 18071 Granada, Spain}%

\author{Airton Deppman}
\email{deppman@usp.br}
\affiliation{Instituto de F\'{i}sica, Universidade de São Paulo, Rua do Matão 1371, São Paulo 05508-090, Brazil}

\begin{abstract}
Using the Plastino-Plastino (PP) equation, we calculate transport coefficients of the heavy-quarks traversing inside the quark-gluon plasma, and generalize their relationship with differential energy loss.  The PP equation indicates anomalous diffusion of the probe particles and yields a quasi-exponential stationary distribution obtained also from the nonadditive statistics proposed by C. Tsallis. We estimate energy loss in a nonadditive quark-gluon medium, and calculate the jet-quenching parameter ($\hat{q}$) for the PP dynamics. With the help of the estimate of $\hat{q}$, we calculate the nuclear suppression factor ($R_{\text{AA}}$) of the heavy-quarks passing through a nonadditive quark-gluon plasma using the model proposed by Dokshitzer and Kharzeev. In many a case, the parameters in the analysis are fixed from the experimental results to minimize arbitrariness. There is a good agreement between the theoretical calculation and experimental $R_{\text{AA}}$ data, indicating that fast heavy-quarks may be subjected to anomalous diffusion inside the QGP. 
\end{abstract}

\maketitle
%
\section{Introduction}
Transport of high-energy particles inside the hot and dense quark-gluon plasma (QGP) medium created in high-energy collisions (e.g. Pb-Pb) is an interesting, 
and active field of research. One of the standard ways to study this phenomenon is to utilize the Boltzmann transport equation (BTE), or its approximations (e.g. the Fokker-Planck equation),
that dictate the evolution of a non-equilibrium distribution (say that of the heavy quarks) inside the QGP medium of the light quarks (LQ), and gluons (g).  Important inputs to these transport equations are the transport
coefficients like drag, and diffusion coefficients that are also related to the energy loss inside the medium. Transport equations are solved using these inputs like the transport coefficients, and the solutions are used to establish a connection with the experimental observables like the nuclear suppression factor.

Transport coefficients are also important inputs to establish the stationary distributions of the transport equations \cite{rafelskiwaltonprl}. Ref.~\cite{rafelskiwaltonprl} shows that using the perturbative QCD estimates of the drag, and the diffusion coefficients of the heavy quarks (HQ), the stationary solution of the Fokker-Planck equation (that the authors initially assumed to belong to a broad class characterized by some parameters) is a quasi-exponential distribution.  On the other hand, it was also shown that, without assuming a mathematical form of the stationary distribution, it is possible to generalize the Boltzmann transport equation \cite{lavagnopla,wilkosada,biroepja} so that the same quasi-exponential function is obtained as the stationary solution in a more straightforward way. This generalized Boltzmann transport equation has been a subject matter in many other works, and the quasi-exponential stationary distribution it yields also arises from the nonadditive statistics (also known as the nonextensive statistics) proposed by C. Tsallis \cite{Tsal88}. 

The nonadditive statistics arises in systems with a fluctuating ambiance \cite{Wilk00,Wilk09}. Imprints of nonadditivity are found in the experimental data obtained in high-energy collision experiments, for example, where the particle
spectra can be described in terms of the following quasi-exponential function \cite{jpg20,Cleymansplb},
\bea
\left(1+\frac{q-1}{T}E\right)^{-\frac{1}{q-1}},
\label{tsallisdist}
\eea
that owes its origin to the Tsallis statistics and converges to the exponential Boltzmann-Gibbs distribution in the limit $q\rightarrow1$. In the above equation $E=\sqrt{P^2+m^2}$ ($P\equiv|\vec{P}|$) is the single particle energy of a particle of the mass $m$ moving with the three-momentum $\vec{P}$, $q$ is the entropic parameter, and $T$ is the temperature parameter. It can be shown that the $q$ parameter is connected to the relative variance in (inverse) temperature \cite{Wilk00}. Also, scaling properties of the Yang-Mills theory shows that $q$ can be deduced from the field theory parameters like $N_c$ (no. of colours) and $N_f$ (no. of flavours) \cite{deppmanprdq}. The $q$ parameter is also utilized to generalize the molecular chaos hypothesis (stosszahlansatz) used in the conventional Boltzmann transport equation. This generalized stosszahlansatz, that may also be a requirement for systems with small number of particles \cite{deppmanqideal}, appears in the generalized Boltzmann transport equation (gBTE) for a non-equilibrium distribution $f$ given below (see, e.g., \cite{wilkosada}),
\bea
P^{\mu} \partial_{\mu} f^q= \mathcal{C}_q[f]~~(\mu=0,1,2,3).
\label{NEBTE}
\eea
It is noteworthy that the power index $q$ on the left hand side ensures energy-momentum conservation \cite{wilkosada}, and the generalized molecular chaos hypothesis is convoluted inside the collision term $\mathcal{C}_q$ that dictates the evolution of the distribution $f$. It is also possible to establish a connection between the quantity $\partial_{\mu}f^q$ and the fractal derivatives. 
In the fractal approach, the Boltzmann equation,
\begin{equation}
  \frac{\partial f}{\partial t}+{\bf v} \cdot \frac{\partial f}{\partial {\bf r}}+ {\bf F} \cdot \frac{\partial f}{\partial {\bf v}}= C[{\bf p}(t+dt),{\bf p}(t)] ,
  \label{eq:BE}
\end{equation}
is modified to incorporate the fractal momentum space by imposing \cite{deppmanfractalfpe}
\begin{equation}
 C[{\bf p}(t+dt),{\bf p}(t)] = D_{t}^{\alpha_q}f \,,
\end{equation}
where $D_{t}^{\alpha_q}f$ is the fractal derivative defined by Parvate and Gangal \cite{parvatefractal}.
%
An appropriate continuous approximation of the fractal derivative leads to the $q$-deformed derivative \cite{deppmanplb,deppmanfractal2}, 
\begin{equation}
 D_{t}^{\alpha_q}f= f^{q-1} \frac{d f}{d t}\,,
\end{equation}
where $\alpha_q-1=1-q$. 
The relaxation time approximation of the BTE 
yields
\begin{equation}
  \frac{\partial f}{\partial t}+{\bf v} \cdot \frac{\partial f}{\partial {\bf r}}+ {\bf F} \cdot \frac{\partial f}{\partial {\bf v}}= -\frac{f-f_0}{\tau} \,,
  \label{eq:relax}
\end{equation}
where $\tau$ is the relaxation time and $f_0$ is the stationary state distribution. The approximation is valid for states near the stationary one.
The term in the r.h.s. of the equation above is a finite difference approximation of the derivative $df/dt$ near the stationary state. By considering the time-momentum space is fractal, this derivative must be substituted by the fractal derivative operator or, in the appropriate continuous approximation, the substitution
\begin{equation}
 \frac{d f_q}{d t} \rightarrow f_q^{q-1} \frac{d f_q}{d t} \sim \frac{d f_q^q}{dt} \,,
 \label{eq:continuousapprox}
\end{equation}
where $f_q$ is a distribution characterized by the $q$ parameter. Considering that the l.h.s. of Eq.~(\ref{eq:relax}) is the time derivative of the distribution $f$, we get
\begin{equation}
 \frac{d f}{d t}= \frac{d f_q^q}{dt}\,. \label{eq:fractalrelax}
\end{equation}
A general solution for this equation can be search with the ansatz
\begin{equation}
 f({\bf p},t)=\sum_{q^{\prime}} f_{q^{\prime}}({\bf p},t)\,.
\end{equation}
However, for systems associated with Tsallis Statistics, it is usually found that the system is characterized by a single and well defined value $q$, so $f \rightarrow f_q$. Due to Eq.~(\ref{eq:relax}), this condition is satisfied only if $f=f_q^q$. 
With this result, Eq.~(\ref{eq:relax}) needs to be rewritten as
\begin{equation}
  \frac{\partial f_q^q}{\partial t}+{\bf v} \cdot \frac{\partial f_q^q}{\partial {\bf r}}+ {\bf F} \cdot \frac{\partial f_q^q}{\partial {\bf v}}= -\frac{f_q-f_0}{\tau} \,,
  \label{eq:relaxq}
\end{equation}
which is the equation utilized in Ref.~\cite{tbphysica}.

The Boltzmann transport equation can be subjected to different approximations like the relaxation time approximation considered above. Recently an iterative scheme has been developed to obtain approximate analytical solutions of the generalized Boltzmann transport equation \cite{tbphysica}. Although the relaxation time approximation of the collision term is a useful first approximation for studying transport phenomena, its assumptions may be valid only for restrictive scenarios \cite{ashcroftbook}. So, attempts have been made to go beyond the relaxation time approximation \cite{maciejprd}. One may also utilize the Landau kinetic approximation \cite{landaubook} that observes that the most of the quark-gluon scattering is soft. In this approximation, it is possible to obtain the Fokker-Planck (FP) equation, and the Plastino-Plastino (PP) equation from the BTE, and the gBTE respectively \cite{tbphysica,svetitsky,deppmanplb}. 

In this work, we investigate the energy loss of the heavy quarks in the quark-gluon plasma medium dictated by the Plastino-Plastino dynamics involving anomalous diffusion of the heavy-quarks. Following are the highlights of our paper: 
\begin{enumerate}
\item We evaluate the transport coefficients from the Plastino-Plastino equation that arises from the generalized Boltzmann transport equation, and has a stationary distribution also obtained from the nonadditive statistics. 
\item We generalize the Rafelski-Walton formula \cite{rafelskiwaltonprl} connecting differential energy loss (-$dE/dr$), drag, and diffusion coefficients, and estimate energy loss per unit length by the heavy quarks traversing through a nonadditive QGP medium. 
\item Later, we compute the jet quenching transport coefficient that is connected to the transverse diffusion coefficient, and provide an estimate of the nuclear suppression factor using the Dokshitzer-Kharzeev (DK) formula \cite{DK} and compare the results with experimental data. Using the DK formula, we also provide analytical results connecting nuclear suppression factor ($R_{\text{AA}}$), jet quenching parameter ($\hat{q}$), and differential energy loss (-$dE/dr$). 
\end{enumerate}

A comparative study of the Fokker-Planck, and the Plastino-Plastino dynamics of the heavy quarks have been performed in a recent article \cite{megiasplb}. However, in this work transport coefficients are estimated following Ref.~\cite{svetitsky} for small incoming momentum of the heavy quarks, and the present work generalizes those estimates. Some estimates of the nuclear suppression factor with the nonextensive initial distributions have been the subject matter in some recent works \cite{tbphysica,tsallisraaepja} that consider the relaxation time approximation. The present work generalizes that framework and the estimates by considering the microscopic scenario of the parton-parton interaction. 

The paper is organized as follows: in the next section we provide the formalism to estimate differential energy loss of the heavy quarks from the Plastino-Plastino (PP) transport coefficients. In Section \ref{qhatandraa}, we compute the jet quenching transport coefficient ($\hat{q}$), and provide an estimate of the nuclear suppression factor in 
the Plastino-Plastino framework. Section \ref{raadedr} contains analytical results establishing a relationship among nuclear suppression factor ($R_{\text{AA}}$), jet quenching parameter ($\hat{q}$), and differential energy loss. Section \ref{discussion} contains a summary of our work, conclusions, and discussion.

\section{Differential energy loss, and the Plastino-Plastino transport coefficients}
\label{energyloss}

\subsection{PP drag and diffusion transport coefficients}
In this section, we shall use an extension of the Rafelski-Walton relationship connecting the transport coefficients, and energy loss. However, before that, we review
how the transport coefficients like drag, and diffusion can be evaluated from the Plastino-Plastino equation. First, we expand the l.h.s. of the gBTE given by Eq.~\eqref{NEBTE} assuming absence of an external force, and obtain the following,
\bea
\frac{\partial f^q(\vec{P},\vec{r},t)}{\partial t} + \vec{v}.\vec{\nabla} f^q(\vec{P},\vec{r},t) = E^{-1}\mathcal{C}_q[f(\vec{P},\vec{r},t)],
\label{NEBTEexpand}
\eea
where $f$ is a non-equilibrium distribution (of the heavy-quarks) dependent on three-momentum $\vec{P} \equiv \{\vec{P_{\text{T}}},P_{\text{L}}\}$ (T: transverse; L: longitudinal), position $\vec{r}$, and time $t$. For a relativistic system, energy of the heavy-quark of the mass $m$ is given by, $E=\sqrt{P^2+m^2}$, and its velocity is given by $\vec{v}=E^{-1}\vec{P}$. 

Now, we assume that the scale of spatial variations along the longitudinal direction is sufficiently large, and the initial motion is one-dimensional along the $x$-direction. By using the Bjorken's assumption of longitudinal boost-invariance (at a time $t$, conditions at point $x$ is the same as those at point $x=0$) at the central rapidity region \cite{baym,japrd}, $f$ obeys
\bea
f(\vec{P}_t,P_x,x,t)=f(\vec{P}_t,P_x',t),
\eea
where $P_x' = (P_x - u_x E)\gamma,~u_x=x/t, $ and $~\gamma=(1-u_x^2)^{-1/2}$. Noting that at $x=0$, $\partial P_x'/\partial x = -E/t$, we find,
\bea
v_x \frac{\partial f^q}{\partial x} = - \frac{P_x}{t} \frac{\partial f^q}{\partial P_x}.
\eea 
Hence, it is possible to write the gBTE for the heavy-quark distributions in the following form \cite{tbphysica},
\bea
\left(\frac{\partial}{\partial t} -\frac{P_x}{t} \frac{\partial}{\partial P_x} \right)f^q(\vec{P}_t,P_x,t)= E^{-1}\mathcal{C}_q[f(\vec{P}_t,P_x,t)].
\label{NEBTEBj}
\eea

For the binary collisions of particles with four-momenta $P^{\mu}\equiv(E,\vec{P})$ (having the momentum distribution $f$), and $Q^{\mu}\equiv(\mathcal{E},\vec{Q})$ (having the momentum distribution $g$), that change to $P'^{\mu}\equiv(E',\vec{P'})$, and $Q'^{\mu}\equiv(\mathcal{E'},\vec{Q'})$ after scattering, the collision term of the gBTE can be written as \cite{wilkosada,biroepja},
\bea
\mathcal{C}_q
&=& \frac{1}{2} \int \frac{d^3\vec{Q}}{(2\pi)^32\mathcal{E}} \frac{d^3\vec{Q'}} {(2\pi)^32\mathcal{E'}} \frac{d^3\vec{P'}} {(2\pi)^3 2E'}  |\overline{\mathcal{M}}|^2 (2\pi)^4 
\delta^4(P+Q-P^{'}-Q^{'}) 
\left[ h_q \{f(\vec{P'}),g(\vec{Q'})\} -h_q \{f(\vec{P}),g(\vec{Q})\} \right].
\label{necollterm} \nn\\
\eea
$|\overline{\mathcal{M}}|^2$ is the square-averaged amplitude of the heavy quark collisional processes (with the light quarks, and gluons) in a quark-gluon plasma medium given by 
Ref.~\cite{combridge}. The function $h_q$ provides an ansatz for the generalization of the molecular chaos hypothesis given by,
\bea
h_q\{f,g\} = \exp_q \left[\log_q (f)+\log_q(g)\right],
\label{hqdef}
\eea
where
\bea
\log_q f = \frac{1-f^{1-q}}{q-1}.
\eea
Here $\log_q$ is the $q$-logarithm function and becomes the conventional logarithm in the limit $q\rightarrow1$,
which also implies, 
\bea
\lim_{q\rightarrow1}h_q\{f,g\} &=& \lim_{q\rightarrow1} \exp_q \left[\log_q (f)+\log_q(g)\right] \nn\\
&=& \exp \left[\log (f)+\log (g)\right] \nn\\
&=& f\times g,
\label{hqdef}
\eea
{\it i.e.}, the distribution functions are independent, which is a consequence of the molecular chaos hypothesis. In this limit, one gets back the
collision term of the conventional BTE \cite{svetitsky},
\bea
\mathcal{C}
&=& \frac{1}{2} \int \frac{d^3\vec{Q}}{(2\pi)^32\mathcal{E}} \frac{d^3\vec{Q'}} {(2\pi)^32\mathcal{E'}} \frac{d^3\vec{P'}} {(2\pi)^3 2E'}  |\overline{\mathcal{M}}|^2 (2\pi)^4 
\delta^4(P+Q-P^{'}-Q^{'}) 
\left[ f(\vec{P'}) g(\vec{Q'}) - f(\vec{P})g(\vec{Q}) \right].
\label{ecollterm} \nn\\
\eea

Using the Landau kinetic approximation, it is possible to obtain \cite{tbphysica} the Plastino-Plastino equation from Eq.~\eqref{NEBTEBj}:
\bea
\left.\frac{\partial f}{\partial t}\right|_{P_xt}= - \frac{\partial}{ \partial P_i} \left(A_{i}^{\text{PP}} f\right) 
+ \frac{\partial}{ \partial P_i \partial P_j} \left(B_{ij}^{\text{PP}} f^{2-q} \right)
~~~~~~~~(i,j=1,2,3).
\label{nefpeIto}
\eea
$B_{ij}^{\text{PP}}$, and $A_{i}^{\text{PP}}$, the Plastino-Plastino diffusion and drag coefficients, are given by
\begin{subequations}
\bea
B_{ij}^{\text{PP}} &=& \frac{f^{q-1}}{2} \left[ \frac{1}{2Eq} 
\int \frac{d^3\vec{Q}}{(2\pi)^32\mathcal{E}} \frac{d^3\vec{Q'}} {(2\pi)^32\mathcal{E'}} \frac{d^3\vec{P'}} {(2\pi)^3 2E'}
~|\overline{\mathcal{M}}|^2 (2\pi)^4 
\delta^4(P+Q-P^{'}-Q^{'}) f^{1-q} S_{\vec{P},\vec{Q}} K_i K_j \right], \nn\\
&&\equiv \frac{f^{q-1}}{2}\langle\langle K_i K_j \rangle\rangle^{PP}
\label{ppdiff}
\eea
\bea
A_{i}^{\text{PP}} &=& \frac{1}{2Eq} 
\int \frac{d^3\vec{Q}}{(2\pi)^32\mathcal{E}} \frac{d^3\vec{Q'}} {(2\pi)^32\mathcal{E'}} \frac{d^3\vec{P'}} {(2\pi)^3 2E'} 
~ |\overline{\mathcal{M}}|^2 (2\pi)^4 
\delta^4(P+Q-P^{'}-Q^{'}) f^{1-q}S_{\vec{P},\vec{Q}} K_i \equiv \left<\left<K_i\right>\right>^{PP}, \nn\\
\label{ppdrag}
\eea
\end{subequations}
where the 3-momentum transfer is $K_i \equiv (P-P')_i$ and 
the factor $S_{\vec{P},\vec{Q}}$, given by
\bea
S_{\vec{P},\vec{Q}} = \left[ 1+ \frac{ \left(1+\delta q_b \frac{|\vec{Q}|}{T_b} \right)^{\frac{\delta q}{\delta q_b}} }
{\left(1+\delta q_h \frac{E}{T}\right)^{\frac{\delta q}{\delta q_h}} }
-\frac{1}{\left(1+\delta q_h \frac{E}{T}\right)^{\frac{\delta q}{\delta q_h}}}\right]^{-\frac{1}{\delta q}},
\label{spq}
\eea
represents the interplay (characterized by $\delta q\equiv q-1$) between the incoming heavy quark distribution (temperature $T$, entropic parameter $\delta q_h\equiv q_h-1$), and the medium distribution (temperature $T_b$, entropic parameter $\delta q_b\equiv q_b-1$). Both the distributions are parameterized by a function similar to Eq.~\eqref{tsallisdist} and for all the calculations in the paper we use $q_h=q_b=1.10,~q=1.002,~T_h=0.4$ GeV, and $T_b=0.2$ GeV. 

The factor $S_{\vec{P},\vec{Q}}$ arises due to generalized molecular chaos in the collision term, and converges to the exponential Maxwell-Boltzmann medium distribution (hence giving rise to molecular chaos) when $q,~q_h$, and $q_b$ approach 1. 


\subsection{Differential energy loss}
\begin{figure}[htbp]
\hspace{-20pt}
 {\includegraphics[width=0.45\textwidth]{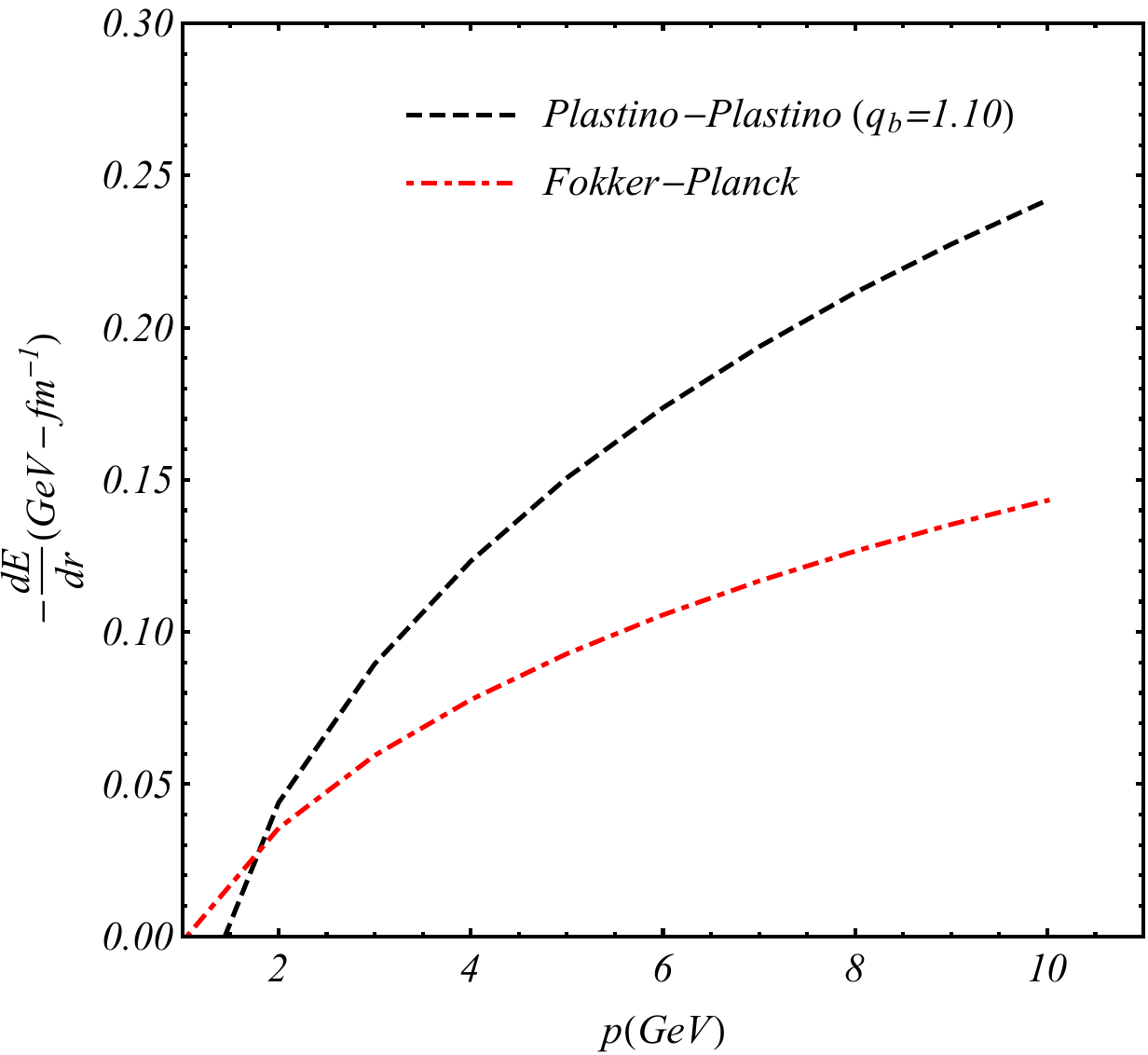}}
\caption{Comparison of the energy loss of the charm quarks ($m$=1.5 GeV) following the FP and PP dynamics in the QGP of $T_b$= 0.2 GeV.}
\label{dedrfig}
\end{figure}

Utilizing the notation used in Eq.~\eqref{ppdiff}, differential energy loss of the heavy quarks in an environment dictated by the non-additive statistics can be written as below following Ref.~\cite{thomaprd}:
\bea
-\frac{dE}{dr} &=& \left \langle \hspace{-4pt}\left\langle \frac{E-E'}{v} \right \rangle \hspace{-4pt} \right\rangle^{PP}
= \left \langle \hspace{-4pt}\left\langle \frac{E^2-EE'}{P} \right \rangle \hspace{-4pt} \right\rangle^{PP}.
\eea
By defining $2B_{00}^{\text{PP}} \equiv f^{q-1} \langle \langle (E-E')^2 \rangle \rangle^{PP}$, and from Eqs.~\eqref{ppdiff}, and \eqref{ppdrag} it is possible to establish a relationship between $-dE/dr$, and the heavy quark transport coefficients as shown below:
\bea
-\frac{dE}{dr} &=& \left \langle \hspace{-4pt}\left\langle \frac{E^2-EE'}{P} \right\rangle \hspace{-4pt} \right\rangle^{PP} \nn\\
&=& \left \langle \hspace{-4pt}\left\langle \frac{P^2+m^2-EE'-\vec{P} \cdot \vec{P'}+\vec{P} \cdot \vec{P'}}{P} \right\rangle \hspace{-4pt} \right\rangle^{PP} \nn\\
&=& \left \langle \hspace{-4pt}\left\langle \frac{1}{P} \left[P^2-\vec{P}.\vec{P'} + \frac{1}{2}\left\{E^2-P^2+E'^2-P'^2-2EE'+2\vec{P} \cdot \vec{P'}\right\} \right] \right\rangle \hspace{-4pt} \right\rangle^{PP} \nn\\
&=& \left \langle \hspace{-4pt}\left\langle \left[ \frac{P_iK_i}{P}  + \frac{1}{2P}\left\{ (E-E')^2-(\vec{P}-\vec{P'})^2\right\} \right] \right\rangle \hspace{-4pt} \right\rangle^{PP}  \nn\\
&=& \frac{P_i}{P}  \left \langle \hspace{-1pt} \left\langle K_i \right\rangle \hspace{-1pt}  \right\rangle^{PP}
+ \frac{1}{2P} \left \langle \hspace{-1pt} \left\langle (E-E')^2 \right\rangle \hspace{-1pt}  \right\rangle^{PP}
- \frac{1}{2P} \left \langle \hspace{-3pt} \left\langle (\vec{P}-\vec{P'})^2 \right\rangle \hspace{-3pt}  \right\rangle^{PP}
\nn\\
&=& \frac{P_iA_{i}^{\text{PP}}+f^{1-q}B_{00}^{\text{PP}}-f^{1-q}B_{ii}^{\text{PP}}}{P}.
\label{dedr}
\eea
In the above equation (also in Eqs.~\ref{bperp}, and \ref{dedrqhat}), repeated indices are summed over. It is noteworthy that in the limit $q\rightarrow 1$, Eq.~\eqref{dedr} converges to a similar equation obtained in Ref.~\cite{rafelskiwaltonprl} for the Boltzmann-Gibbs statistics. In the high-energy regime ($E,E'>>m$) where small angle scattering is more dominant, the above equation can be approximated as,
\bea
-\frac{dE}{dr} &\approx& \gamma^{\text{PP}} P,
\eea
where $A_{i}^{\text{PP}}=\gamma^{\text{PP}} P_i$, and $P_iP_i=P^2$. We evaluate the transport coefficients following the method elaborated in Ref.~\cite{svetitsky}. Fig.~\ref{dedrfig} compares the differential energy loss in the Plastino-Plastino, and the Fokker-Planck
formalism. It shows that the Plastino-Plastino energy loss starts surpassing the Fokker-Planck counterpart at around 2 GeV momentum of the incoming charm quark. 

Experimental evidences of energy loss in the QGP is manifest through measuring the nuclear suppression factor defined by,
\be
R_{\mathrm{AA}} = \frac{\left(d^2N/dp_{\mathrm{T}}dy\right)^{\mathrm{A+A}}}
{N_{\mathrm{coll}}\times\left(d^2N/dp_{\mathrm{T}}dy\right)^{\mathrm{p+p}}}.
\ee
$N_{\mathrm{coll}}$ is the number of nucleon-nucleon binary collisions (p+p) while a nucleus `A' collides with another nucleus. $d^2N/dp_{\mathrm{T}}dy$
denotes the differential yield within the transverse momentum and rapidity range $p_{\rm{T}}$ to $p_{\mathrm{T}}+dp_{\mathrm{T}}$ and $y$ to $y+dy$. 
Theoretical estimates of $R_{\mathrm{AA}}$ are obtained by solving the transport equation. However, in this work, we aim to estimate $R_{\text{AA}}$ by computing another transport coefficient $\hat{q}$, the jet-quenching parameter, that yields the transverse momentum broadening per unit length. Dokshitzer and Kharzeev \cite{DK} have derived an analytical result connecting $\hat{q}$, and $R_{\mathrm{AA}}$, and the rest of the article will be devoted to computing $\hat{q}$, and providing numerical estimates of $R_{\mathrm{AA}}$. We also will discuss analytical results connecting the transport coefficients and nuclear suppression factor.

\section{Calculating the jet-quenching parameter $\hat{q}$}
\label{qhatandraa}


In the previous section, we have derived the Plastino-Plastino equation from the gBTE, and found the integral expressions for the drag and diffusion coefficients, 
$A^{\text{PP}}_{i}$, and $B^{\text{PP}}_{ij}$. The diffusion tensor $B^{\text{PP}}_{ij}$ can be written in terms of a longitudinal and a transverse diffusion coefficient  
in the following way \cite{svetitsky},
\bea
B^{\text{PP}}_{ij} &=& \left(\delta_{ij}-\frac{{P_i}{P_j}}{{{P}^2}}\right) B^{\text{PP}}_{\text{T}} ({P^2})+ 
\frac{{P_i}{P_j}}{{P^2}} B^{\text{PP}}_{\text{L}} (P^2) ~~~(P^2 \equiv |\vec{P}|^2).
\label{bij}
\eea
Hence, the trace of the diffusion tensor (that appears in Eq.~\ref{dedr}) is $B_{ii}^{\text{PP}}=2B^{\text{PP}}_{\text{T}}+B^{\text{PP}}_{\text{L}}$.
From Eq.~\eqref{bij} we obtain,
\bea
 B^{\text{PP}}_{\text{T}} &=& \frac{1}{2} \left(\delta_{ij}-\frac{{P_i}{P_j}}{{P^2}}\right) B^{\text{PP}}_{ij} \nn\\
 &=& \frac{f^{q-1}}{4} \left(\delta_{ij}-\frac{{P_i}{P_j}}{{P^2}}\right) \langle\langle K_i K_j \rangle\rangle^{PP} \nn\\
 &=& \frac{f^{q-1}}{4} \left\langle \hspace{-4pt} \left\langle K^2 -\frac{(\vec{P} \cdot \vec{K})^2}{P^2} \right\rangle \hspace{-4pt} \right\rangle^{PP} \nn\\
 &=& \frac{f^{q-1}}{4} \left\langle \hspace{-2pt} \left\langle K_{\text{T}}^2 \right\rangle \hspace{-2pt} \right\rangle^{PP},
 \label{bperp}
\eea
where $\vec{K} \equiv (\vec{K}_\text{T},K_{\text{L}}) \equiv (\vec{K}_\text{T}, \vec{K} \cdot \vec{P}/P) $. 
Further, $\langle\langle K_{\text{T}}^2 \rangle\rangle$, that gives transverse momentum broadening per unit time, can be connected to the jet-quenching coefficients $\hat{q}$ that estimates average transverse momentum broadening per unit length in the following way \cite{SolovevaPRD}:
\bea
\hat{q} = \frac{ \langle \hspace{-1pt} \langle K_{\text{T}}^2 \rangle \hspace{-1pt} \rangle}{v_\text{L}}= \frac{E}{P_\text{L}} \langle \hspace{-1pt} \langle K_{\text{T}}^2 \rangle \hspace{-1pt} \rangle ~~(v_\text{L}:~\text{longitudinal~velocity}).
\eea
Hence, from Eq.~\eqref{bperp}
\bea
\hat{q}^{\text{PP}}= 4f^{1-q}\frac{E}{P_\text{L}} B^{\text{PP}}_{\text{T}}.
\label{qhat}
\eea
It is noteworthy that, in the limit $q\rightarrow1$, a similar relationship between $\hat{q}$, and $B_{\text{T}}$ has also been derived in Ref.~\cite{TBPrd} for the kinematic region $P_\text{L}>>m,P_{\text{T}}$ that implies $\hat{q} \approx 4 B_{\text{T}}$.


Now, the jet-quenching transport coefficient $\hat{q}$ is connected to the nuclear suppression factor $R_{\text{AA}}$ that is an experimental observable. So, with our estimates of $\hat{q}$, we are also able to provide an estimate of $R_{\text{AA}}$ in a nonadditive medium. In particular, it will be interesting to see what effect 
the entropic $q$ parameter (that parameterizes the generalized molecular chaos) has on heavy-quark suppression.
An analytical formula relating $\hat{q}$ and the nuclear
suppression factor $R^{\text{h}}_{\text{AA}}$ for the heavy quarks (denoted by `h') has been used in Ref.~\cite{laceyPRL}, and derived in Ref.~\cite{DK}. This formula is given by (for a path length $\ell$, strong coupling $\alpha_s$, and colour factor $C_F$),

\bea
R^{\text{h}}_{\text{AA}}(P_{\text{T}},\ell) \simeq 
\exp 
\left[ -\frac{2\alpha_s C_F}{\sqrt{\pi}} 
\ell \sqrt{\frac{\hat{q} \mathcal{L}_h^{\text{abs}}}{P_{\text{T}}}} + \frac{16\alpha_s C_F}{9\sqrt{3}} \ell \left( \frac{\hat{q}m^2}{m^2+P_{\text{T}}^2} \right)^{1/3} \right].
\label{DKformula}
\eea
However, this formula for $R^{\text{h}}_{\text{AA}}$ is derived considering radiative processes. Hence, we need to calculate the radiative transport coefficients to be discussed in the following section, along with the evaluation of the quantity $\mathcal{L}_h^{\text{abs}}$.

\label{calcraa}


\subsection{Estimating the jet quenching parameter $\hat{q}$: collisional and radiative}
\label{evalqhat}
\begin{figure}[htbp]
\begin{center}
 {\includegraphics[width=0.40\textwidth]{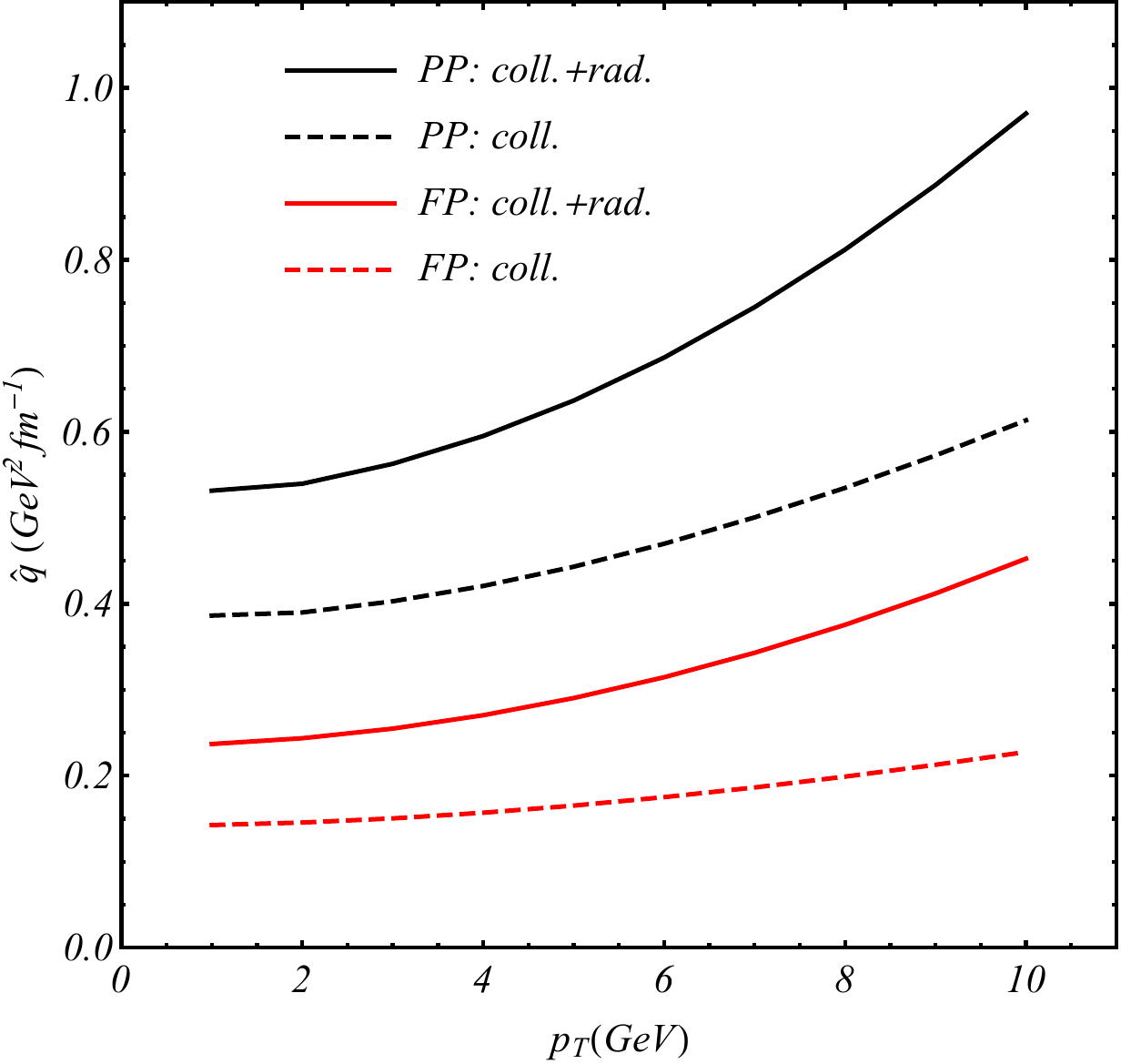}}
\caption{Comparison of total (collisional+radiative), and collisional jet-quenching parameters ($\hat{q}$)  of the charm quarks ($m$=1.5 GeV; $P_{\text{L}}$=10 GeV) following the FP and PP ($q_b$=1.10) dynamics in the QGP of $T_b$= 0.2 GeV.}
\label{qhatpt}
\end{center}
\end{figure}
According to Eq.~\eqref{qhat}, $\hat{q} = 4 (E/P_\text{L}) f^{1-q}B_{\text{T}}$, and is given by the following integration in the PP case for the collisional (e.g. HQ+LQ/g $\rightarrow$ HQ+LQ/g) processes.
\bea
\hat{q}^{\text{PP}}_{\text{coll}} = \frac{1}{2qP_\text{L}} 
\int \frac{d^3\vec{Q}}{(2\pi)^3 2\mathcal{E}} \frac{d^3\vec{Q'}} {(2\pi)^3 2\mathcal{E'}} \frac{d^3\vec{P'}} {(2\pi)^3 2E'} 
|\overline{\mathcal{M}}|_{2\rightarrow2}^2 (2\pi)^4 \delta^4(P+Q-P'-Q') S_{\vec{P},\vec{Q}} \left(P'^2 -\frac{(\vec{P}.\vec{P'})^2}{P^2} \right),
\eea

For the radiative processes (e.g. HQ+LQ/g $\rightarrow$ HQ+LQ/g+g), we follow Refs.~\cite{TBPrd,SKDPrd} to calculate the radiative transport coefficients from the 
collisional ones. To be consistent with Eq.~\eqref{DKformula}, we consider only collinear, and soft gluon (4-momentum $k_5^{\mu}\equiv \{E_5,\vec{k}_5\}$) radiation 
without the Bose enhancement factor in the present calculation. These considerations lead to the
following estimate of the radiative jet transport coefficient,
\bea
\hat{q}^{\text{PP}}_{\text{rad}} &=& \frac{1}{2qP_\text{L}} \int \frac{d^3\vec{Q}}{(2\pi)^3 2\mathcal{E}} \frac{d^3\vec{Q'}} {(2\pi)^3 2\mathcal{E'}} \frac{d^3\vec{P'}} {(2\pi)^3 2E'} \frac{d^3 \vec{k_5}}{(2\pi)^3 2E_5}
(2\pi)^4 \delta^4(P+Q-P'-Q'-k_5) S_{\vec{P},\vec{Q}} \left(P'^2 -
\frac{(\vec{P}.\vec{P'})^2}{P^2} \right) \nn\\
&&\times ~|\overline{\mathcal{M}}|_{2\rightarrow3}^2 \theta_1\left(E-E_5\right) \theta_2\left(\tau-\tau_{\text{F}}\right),
\eea
where the first $\theta$-function prohibits gluon radiation of energy greater than that of the incoming heavy quark, and the second $\theta$-function accounts for the additive region of the gluon spectra due to multiple scattering in which the interaction time ($\tau$) in greater than the formation time ($\tau_{\text{F}}$) of the gluons \cite{PlumerPRD}. These two $\theta$-functions put an upper-bound and a lower-bound respectively in the gluon 3-momentum 
($\vec{k}_5 \equiv \{\vec{k}_{\text{T}},k_{\text{L}}\}$) integration. The radiative matrix element for the soft, collinear gluon emission at an angle $\theta_g$ can be written in the following way \cite{DK} (for $g_s^2=4\pi\alpha_s$; $\alpha_s\equiv$ strong coupling),
\bea
|\overline{\mathcal{M}}|_{2\rightarrow3}^2 \approx |\overline{\mathcal{M}}|_{2\rightarrow2}^2 \times \frac{12g_s^2}{k_{\text{T}}^2} \left(1+\frac{m^2}{E^2\theta_g^2}\right)^{-2}.
\eea
Also, parameterizing the soft gluon 4-momentum in terms of its rapidity ($\eta$), we get,
\bea
\hat{q}^{\text{PP}}_{\text{rad}} &\approx& \int \text{collisional~part} \times \frac{6\alpha_s}{\pi}  \int d\eta \int \frac{d k_{\text{T}}} {k_{\text{T}}}   ~ \left(1+\frac{m^2}{E^2\theta_g^2}\right)^{-2} \theta_1\left(E-E_5\right) \theta_2\left(\tau-\tau_{\text{F}}\right).
\eea
Now, using the parameterization $E_5=k_{\text{T}} \cosh\eta$, putting $\tau_{\text{F}}=\cosh\eta/k_{\text{T}}$, and using $\theta_g = 2\tan^{-1}\left[\exp(-\eta)\right]$,
\bea
\hat{q}^{\text{PP}}_{\text{rad}} &\approx& \int \text{collisional~part} \times \frac{6\alpha_s}{\pi}  \int d\eta \int_{\Lambda\cosh\eta}^{\frac{E}{\cosh\eta}} \frac{d k_{\text{T}}} {k_{\text{T}}}   ~ \left[ 1+\frac{m^2}{E^2 \left(2\tan^{-1}\left[\exp(-\eta)\right]\right)^2}\right]^{-2}.
\eea
In the above integration, the interaction rate $\Lambda$ is evaluated from the following integral,
\bea
\Lambda = \frac{1}{2qE} \int \frac{d^3\vec{Q}}{(2\pi)^3\mathcal{E}} \frac{d^3\vec{Q'}} {(2\pi)^3\mathcal{E'}} \frac{d^3\vec{P'}} {(2\pi)^3 E'} 
|\overline{\mathcal{M}}|_{2\rightarrow2}^2 (2\pi)^4 \delta^4(P+Q-P'-Q') f^{1-q}S_{\vec{P},\vec{Q}}.
\eea

In Fig.~\ref{qhatpt} we show the variation of the (Plastino-Plastino and Fokker-Planck) jet transport coefficient $\hat{q}^{\text{PP/FP}}=\hat{q}^{\text{PP/FP}}_{\text{coll}} + \hat{q}^{\text{PP/FP}}_{\text{rad}}$ with the incoming HQ transverse momentum $p_{\text{T}}$.

\subsection{Estimating $\mathcal{L}_h^{\text{abs}}$, and $R_{\text{AA}}$ and comparing with experimental data}
\label{lh}

\begin{figure}[h]
\begin{center}
 {\includegraphics[width=0.40\textwidth]{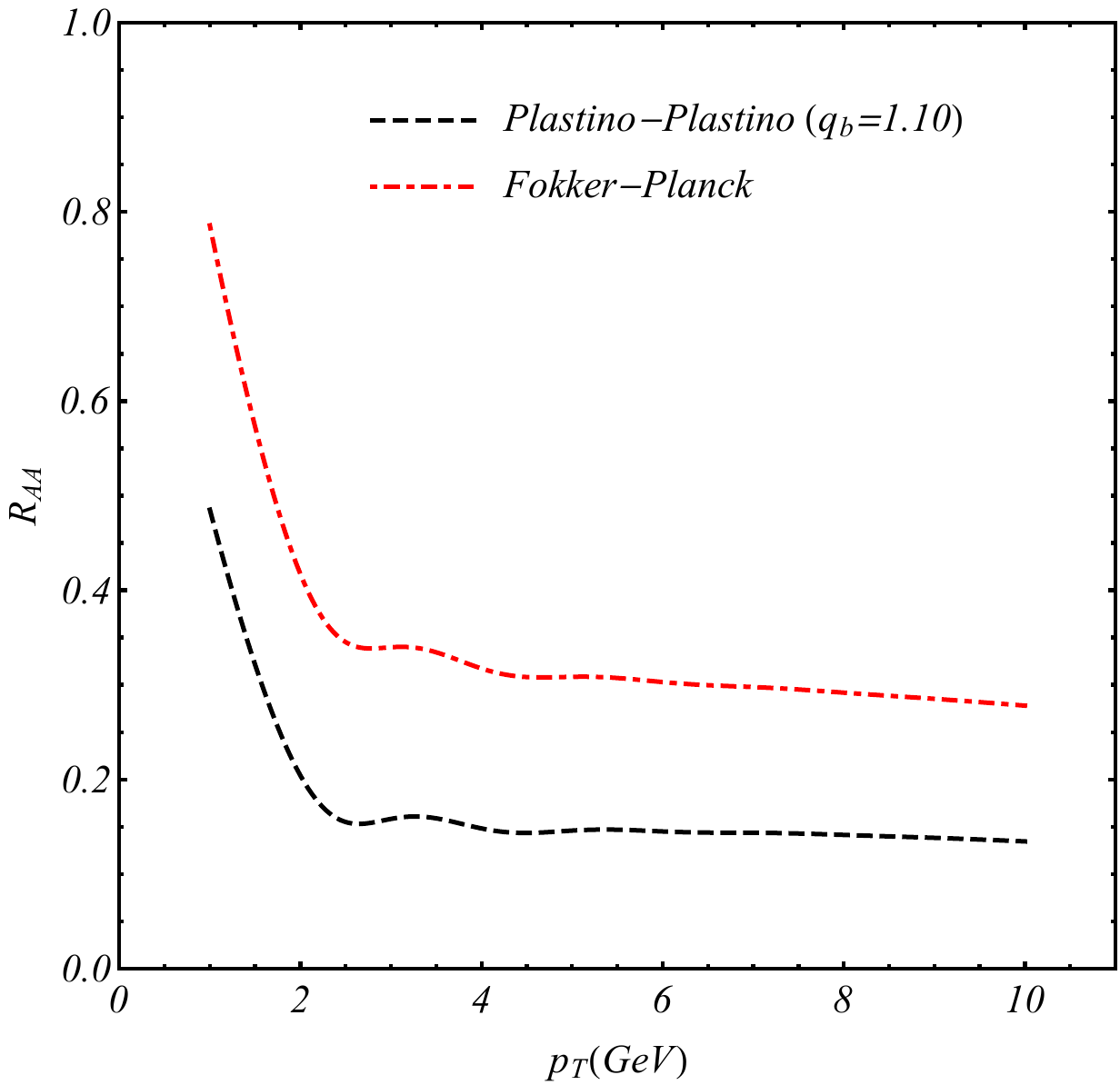}}
\caption{Comparison of the nuclear suppression factor ($R_{\text{AA}}$)  of the charm quarks ($m$=1.5 GeV; $P_{\text{L}}$=10 GeV) following the FP and PP dynamics in the QGP of $T_b$= 0.2 GeV calculated using Eq.~\eqref{DKformula}. We put $b=1.7,~n=11,~\ell=3$ fm.}
\label{RAApt}
\end{center}
\end{figure}
For evaluating $\mathcal{L}_h^{\text{abs}}$, we shall follow Ref.~\cite{DK} that defines 
\bea
\mathcal{L}_h = \frac{d}{d\ln p_{\text{T}}} \ln \left[ \frac{d\sigma^{\text{vac}}} {dp_{\text{T}}^2} \left(p_{\text{T}}\right) \right],
\eea
and uses the following vacuum differential cross-section for the D-mesons \cite{dmeson}
\bea
\frac{ d\sigma^{\text{vac}} } {dp_{\text{T}}^2} = C \left( \frac{1}{bm^2+ p_{\text{T}}^2}  \right)^{\frac{n}{2}},
\eea
for $b=1.4\pm0.3$, $n=10.0\pm1.2$, and $m=1.5$ GeV. Hence, the quantity $\mathcal{L}_h$ can be evaluated to be,
\bea
\mathcal{L}_h = -\frac{n p_{\text{T}}^2}{bm^2+ p_{\text{T}}^2} \Rightarrow \mathcal{L}_h^{\text{abs}} = \left|\mathcal{L}_h\right|.
\eea
We plot the variation of the charm quark nuclear suppression factor with its transverse momentum in Fig.~\ref{RAApt}.

\begin{figure}[h]
\begin{center}
 {\includegraphics[width=0.4\textwidth]{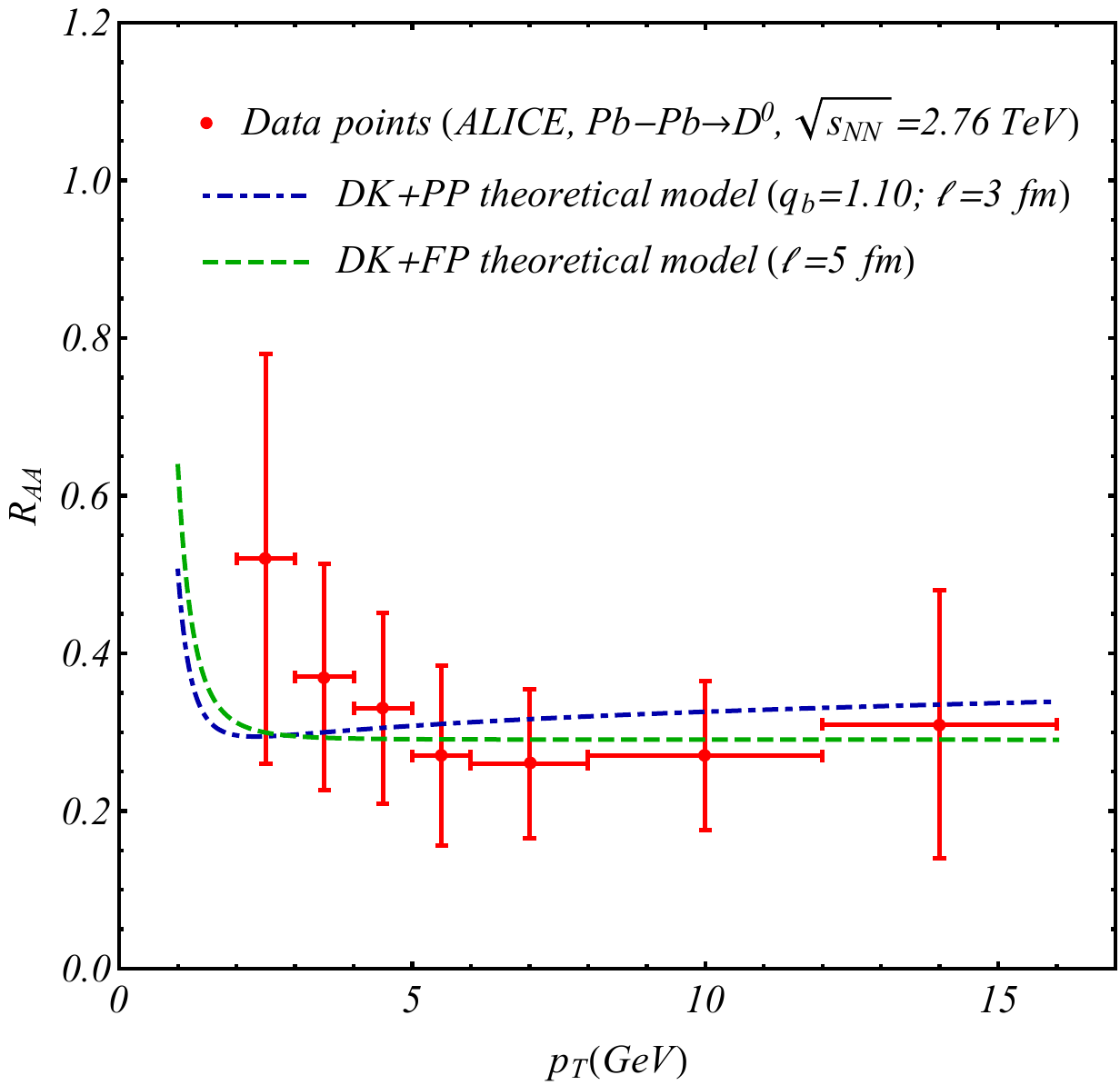}}
\caption{Comparison of nuclear suppression factor of the $D^0$ mesons calculated using Eqs.~\eqref{DKformula}, and \eqref{fragfunc} for the PP and FP dynamics with experimental data obtained by the ALICE collaboration \cite{ALICED0}. We put $T_b=0.2$ GeV, $b$=1.7, $n$=11, and $\epsilon_{\text{h}}$=0.056.}
\label{RAADmesonpt}
\end{center}
\end{figure}

We also compare the theoretical model calculations with the experimental data \cite{ALICED0} of the $D^0$ mesons in Fig.~\ref{RAADmesonpt}. To this end, we integrate Eq.~\eqref{DKformula} with the following function \cite{peterson} fragmenting heavy-quarks (`h') to the hadrons (`H') with the momentum fraction $z$ of the partons:
\bea
D^{\text{H}}_{\text{h}} (z) \propto \frac{1}{z\left(1-\frac{1}{z} - \frac{\epsilon_{\text{h}}}{(1-z)} \right)^2}.
\label{fragfunc}
\eea


\section{Relationship among jet quenching parameter, differential energy loss, and nuclear suppression factor}
\label{raadedr}
From Eqs.~\eqref{dedr}, and \eqref{qhat}, one can establish a relationship between $-dE/dr$, and $\hat{q}$ in the following way:
\bea
-\frac{dE}{dr} &=& \frac{P_iA_{i}^{\text{PP}}+f^{1-q}B_{00}^{\text{PP}}-f^{1-q}B_{ii}^{\text{PP}}}{P} \nn\\
&=& \frac{P_iA_{i}^{\text{PP}}+f^{1-q} (B_{00}^{\text{PP}}-2B^{\text{PP}}_{\text{T}}-B^{\text{PP}}_{\text{L}})}{P} \nn\\
&=& P\gamma^{\text{PP}} + f^{1-q} \left( \frac{B_{00}^{\text{PP}}-B^{\text{PP}}_{\text{L}}}{P} \right) - \frac{\hat{q}^{\text{PP}} P_{\text{L}}}{2PE}
\label{dedrqhat}
\eea
In the Boltzmann-Gibbs (BG) limit (in which case we simply remove the superscript `PP'), the above equation becomes,
\bea
-\frac{dE}{dr} = P\gamma + \frac{B_{00}-B_{\text{L}}}{P} - \frac{\hat{q} P_{\text{L}} }{2PE}.
\label{dedrqhatbg}
\eea
The transport coefficients $\gamma$, $B_{00}$, $\hat{q}$ and $B_{\text{L}}$ are the conventional Fokker-Planck transport coefficients. 
Eq.~\eqref{dedrqhat} also implies,
\bea
\hat{q}^{\text{PP}} &=& \frac{2PE}{P_{\text{L}}} \left[ P\gamma^{\text{PP}} + f^{1-q}\left(\frac{B_{00}^{\text{PP}}-B^{\text{PP}}_{\text{L}}}{P}\right) + \frac{dE}{dr} \right]
\text{(Plastino-Plastino),}~~\text{and} \nn\\
\hat{q} &=& \frac{2PE}{P_{\text{L}}} \left( P\gamma + \frac{B_{00}-B_{\text{L}}}{P} + \frac{dE}{dr} \right)~~~~\text{(Fokker-Planck)}.
\label{qhatdedr}
\eea

From Eq.~\eqref{qhatdedr}, it is possible to establish a connection between nuclear suppression factor, and differential energy loss in the following way:
\bea
R^{\text{h}}_{\text{AA}}(P_{\text{T}},\ell) \simeq 
\exp 
\left[ -\frac{2\alpha_s C_F}{\sqrt{\pi}} 
\ell 
\sqrt{\frac{\frac{2PE}{P_{\text{L}}} \left( P\gamma^{\text{PP}} + \frac{f^{1-q}(B_{00}^{\text{PP}}-B^{\text{PP}}_{\text{L}})}{P} + \frac{dE}{dr} \right) \mathcal{L}_h^{\text{abs}}}{P_{\text{T}}}} \right.\nn\\
\left.+ \frac{16\alpha_s C_F}{9\sqrt{3}} 
\ell 
\left( \frac{\frac{2PE}{P_{\text{L}}} \left( P\gamma^{\text{PP}} + \frac{f^{1-q}(B_{00}^{\text{PP}}-B^{\text{PP}}_{\text{L}})}{P} + \frac{dE}{dr} 
\right)m^2}{m^2+P_{\text{T}}^2} \right)^{1/3} \right],
\label{DKformuladedr}
\eea
that, in the BG limit ($q\rightarrow 1$), becomes,
\bea
R^{\text{h}}_{\text{AA}}(P_{\text{T}},\ell) \simeq 
\exp 
\left[ -\frac{2\alpha_s C_F}{\sqrt{\pi}} 
\ell \sqrt{\frac{\frac{2PE}{P_{\text{L}}} \left( P\gamma + \frac{B_{00}-B_{\text{L}}}{P} + \frac{dE}{dr} \right) \mathcal{L}_h^{\text{abs}}}{P_{\text{T}}}} \right.\nn\\
\left.+ \frac{16\alpha_s C_F}{9\sqrt{3}} \ell \left( \frac{\frac{2PE}{P_{\text{L}}} \left( P\gamma + \frac{B_{00}-B_{\text{L}}}{P} + \frac{dE}{dr} \right)m^2}{m^2+P_{\text{T}}^2} \right)^{1/3} \right].
\label{DKformuladedrbg}
\eea

\section{Discussion and conclusion}
\label{discussion}
In this work, we revisit the derivation of the Plastino-Plastino equation from a generalized Boltzmann transport equation. The index $q$ also appears in Ref.~\cite{deppmanfractalfpe} (as a power of the test particle distribution) that derives the Plastino-Plastino equation using the fractal space approach.
We calculate the transport coefficients like the diffusion and the jet-quenching parameter of the charm quarks, and differential energy loss in a nonadditive QGP medium. We have generalized some previously used relationships (Eqs.~\eqref{dedr}, and \eqref{qhat}), and presented analytical formulae establishing a connection between $R_{\text{AA}}$, an experimental observable, and the transport coefficients using Ref.~\cite{DK}
(Eqs.~\eqref{DKformuladedr}, and \eqref{DKformuladedrbg}). From this paper it is also possible to estimate the elliptic flow parameter by defining path-length-dependent  $R_{\text{AA}}$ for the in-plane and out-of-plane directions \cite{RAAv2}.

We observe that when there is an interplay between the incoming heavy-quarks and the medium particle distribution, manifested by a generalized molecular chaos, numerical values of transport coefficients, energy loss, and suppression of the heavy-quarks increase compared to their Boltzmann-Gibbs counterparts.
Given that for finite systems generalized molecular chaos may be a more realistic scenario \cite{deppmanqideal}, our estimates in this paper will be relevant for the QGP system. The $\hat{q}$ values obtained using the PP dynamics is around 1 GeV$^2$-fm$^{-1}$ which is comparable to the value obtained in Ref.~\cite{laceyPRL} and less compared to the `experimentally favoured range' for the RHIC data \cite{armestoqhat}. However, the DK formula considers soft, eikonal and collinear gluon radiation approximation, and it will be interesting to investigate how generalizing these approximation considered in Ref.~\cite{DK}, following Ref.~\cite{tbnoneikonal}, will impact the estimate of $\hat{q}$, and other experimental observables. With these modifications, it can also be expected that the contribution of radiation in $\hat{q}$ will increase, more so for the PP case where $\hat{q}_{\text{rad}}$ is around 40-60\% of $\hat{q}_{\text{coll}}$. 
A few salient points that come out of the discussion are listed below:
\begin{itemize}
\item Calculation of $R_{\text{AA}}$ is based on calculating the jet transport coefficient evaluated using the Plastino-Plastino equation that takes into account  
anomalous diffusion of the heavy-quarks in the medium.
\item In many a case, the parameters in the analysis are fixed from the experimental results to minimize arbitrariness.
\item For the given parameter values, there is a good agreement between the theoretical calculation and $R_{\text{AA}}$ data, indicating that fast heavy-quarks may be subjected to anomalous diffusion inside the QGP \cite{anodiff}.
\end{itemize}
Our calculation of $R_{\text{AA}}$ using the PP dynamics captures the trend of decreasing suppression (considering the central values) at higher transverse momenta (Fig.~\ref{RAADmesonpt}), but the FP dynamics does not show this trend. It can be intuitively understood that $R_{\text{AA}}$ depends on $\ell$ that may be equalled to the average path length traversed by the heavy-quarks. This quantity $\ell$ can introduce a centrality dependence in our calculations. We observe that for a 0-20 \% central collision between nuclei with the same radius (e.g. Pb on Pb), the average path length of the heavy-quarks varies between 55\% to 75 \% of the radius of the produced system. So, our analysis with the PP dynamics indicates that the diameter of the QGP system produced may be around 8-10 fm, but the FP dynamics indicates a larger diameter of around 13-18 fm. Nevertheless, we think that a more detailed analysis involving centrality dependent experimental data and interferometry may be required. For the lower $p_{\text{T}}$ values, more involved investigation (perhaps involving collectivity) may benefit the study. This also begs a more rigorous approach to establish a connection between theory and experiment by utilizng a numerical solution of the Plastino-Plastino equation with the transport coefficients as inputs. In this extension, the collective behaviour of medium (that is not considered in the present work) should be included for a more reliable comparison with experimental data. We reserve these works for the future.




\section*{Acknowledgements}
The work of EM is supported by the project PID2020-114767GB-I00 and by the Ram\'on y Cajal Program under Grant RYC-2016-20678 funded by MCIN/AEI/10.13039/501100011033 and by ``FSE Investing in your future'', by Junta de Andaluc\'{\i}a under Grant FQM-225, and  by the ``Pr\'orrogas de Contratos Ram\'on y Cajal'' Program of the University of Granada.  This work is a part of the project INCT-FNA Proc. No. 464898/2014-5. A.D. acknowledges the support from the Conselho Nacional de Desenvolvimento Cient\'{\i}fico e Tecnol\'ogico (CNPq-Brazil), grant 306093/2022-7.

\end{document}